\documentclass[12pt,preprint]{aastex}
\begin{document}

\title{Stochastic relativistic shock-surfing acceleration}
\author{Benjamin D. G. Chandran }
\email{benjamin-chandran@uiowa.edu} 
\author{Naoki Bessho\footnote{Present address: Department of Physics \& Astronomy, University of New Hampshire}}
\email{naoki.bessho@unh.edu}
 \affil{Department of Physics \& Astronomy, University of Iowa}
\begin{abstract}
We study relativistic particles undergoing surfing
acceleration at perpendicular shocks. We assume
that particles undergo diffusion in the component
of momentum perpendicular to the shock plane due
to moderate fluctuations in the shock electric and
magnetic fields. We show that $dN/dE$, the number
of surfing-accelerated particles per unit energy,
attains a power-law form,~$dN/dE \propto E^{-b}$.
We calculate~$b$ analytically in the limit of weak
momentum diffusion, and use Monte Carlo
test-particle calculations to evaluate~$b$ in the
weak, moderate, and strong momentum-diffusion
limits.
\end{abstract}
\maketitle

\section{Introduction}

\begin{figure}[h]
\vspace{9cm}
\includegraphics{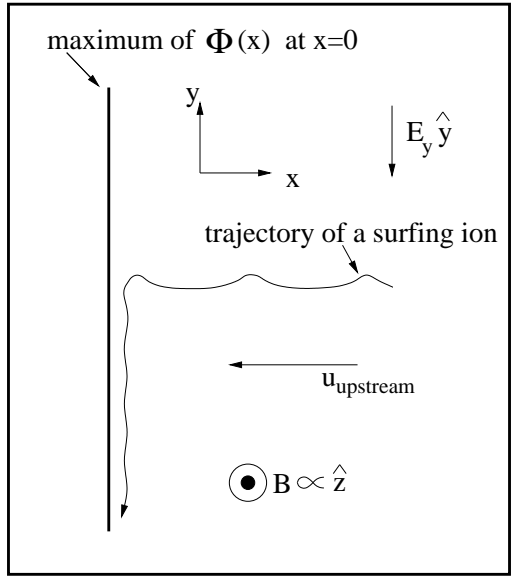}
\caption{Idealized trajectory of a surfing ion at a perpendicular
shock as seen in the shock wave frame.
\label{fig:f1}}
\end{figure}

\begin{figure}[h]
\vspace{9cm}
\includegraphics{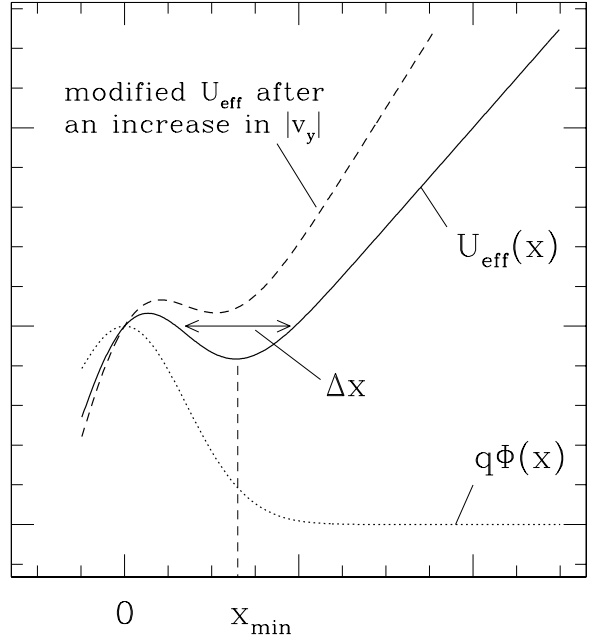}
\caption{Idealized plot of electrostatic potential energy~$q\Phi(x)$
and effective potential energy~$U_{\rm eff}(x)$.
\label{fig:f2}}
\end{figure}

The acceleration of high-energy particles is an important problem in
astrophysics. This paper examines one acceleration mechanism,
``surfing acceleration,''\footnote{See, e.g., Sagdeev (1966), Sagdeev
\& Shapiro (1973), Katsouleas \& Dawson (1983), Ohsawa \& Sakai
(1987), Lee et~al (1996), Zank et~al (1996), Lipatov \& Zank (1999),
Ucer \& Shapiro (2001), McClements et~al (2001), Lever, Quest, \&
Shapiro (2001), Shapiro, Lee, \& Quest (2001), and Hoshino \& Shimada
(2002).}  as it applies to relativistic particles at perpendicular
shocks.  An illustration of an ion undergoing surfing acceleration at
a perpendicular shock is given in figure~\ref{fig:f1}. In this picture
an ion of velocity~${\bf v}$ arrives just upstream of the shock with
small $|v_x|$ and is reflected by the jump in the electrostatic
potential~$\Phi(x)$ at the shock, which is illustrated schematically
in figure~\ref{fig:f2}.  The Lorentz force then causes the ion to
return to the shock where it is again reflected and 
brought back to the shock front by the Lorentz force. This
process continues, confining the particle to the vicinity of the
shock, where it is accelerated in the~$-\hat{{\bf y}}$ direction by the
motional electric field,~$-{\bf u} \times {\bf B} /c$, where~${\bf u}$
is the plasma velocity.  If one thinks of~$q\Phi(x)$ in figure~\ref{fig:f2} 
as analogous to the profile of an ocean wave, the accelerating ion in 
figure~\ref{fig:f1} is analogous to a surfer.
In this paper, we will focus on ions.
However, shock surfing acceleration is believed to be
important for electrons as well (McClements et~al 2001, Hoshino \&
Shimada 2002), and it is trivial to modify the analysis of
sections~\ref{sec:analytic} and~\ref{sec:num} to treat the electron
case.

An ion's $x$-oscillations in front of the shock can be
approximately described in terms of an effective potential energy
(Lee et~al 1996, Ucer \& Shapiro 2001),
which can be derived as follows . We start with the $x$-component of
the equation of motion,
\begin{equation}
m\frac{d}{dt}\left(\gamma v_x\right) = q \left(E_x + \frac{v_y B_z - v_z B_y}{c}\right),
\label{eq:dpxdt} 
\end{equation} 
where~$\gamma$ is the relativistic Lorentz factor, and we work in
shock wave frame.  We assume that $E_x = - d\Phi/dx$ with~$\Phi$ a
function of~$x$ alone, that~$|v_z B_y| \ll |v_y B_z|$, that~$|v_x| \ll
|v_y|$, and that the particle oscillates in~$x$ on a time scale that
is much shorter than the time required for~$\gamma$ or~$v_y$ to change
appreciably.  We also treat~$B_z$ as spatially uniform.  We then
multiply equation~(\ref{eq:dpxdt}) by~$v_x$ and integrate in time,
treating~$\gamma$ and~$v_y$ as constant, to obtain
\begin{equation}
\frac{m\gamma v_x^2}{2} + U_{\rm eff}(x) = \mbox{ constant } \equiv w + U_{\rm min},
\label{eq:xenergy} 
\end{equation} 
where
\begin{equation}
U_{\rm eff}(x) = q\left[\Phi(x) - \frac{x v_y B_z}{c}\right],
\label{eq:ueffdef0} 
\end{equation} 
$w$ is the energy of
$x$-oscillations, and $U_{\rm min}$ is the value of~$U_{\rm eff}$
at its local minimum at~$x=x_{\rm min}$. We define 
\begin{equation}
w_{\rm max} = U_{\rm max} - U_{\rm min},
\label{eq:defwmax0} 
\end{equation} 
where $U_{\rm max}$ is
the value of $U_{\rm eff}(x)$ at its local maximum. Surfing
ions, which are trapped in the effective potential
well, have~$0 < w< w_{\rm max}$.
The effective potential energy is
plotted  in figure~\ref{fig:f2} for~$v_y < 0$. 
The turning points of an ion of given~$w$,
obtained by solving equation~(\ref{eq:xenergy}) with~$v_x=0$,
are denoted~$x_L(w)$ and $x_R(w)$.
Over many bounces, $\gamma$ and~$v_y$ vary slowly, causing~$w$
to change in a way that can be calculated by noting that
the adiabatic invariant
\begin{equation}
J = \int_{x_L(w)}^{x_R(w)} p_x \;dx
\label{eq:defJ0} 
\end{equation} 
is approximately conserved (Ucer \& Shapiro 2001).

Surfing can only occur if
there is a local minimum in~$U_{\rm eff}$, which requires that
\begin{equation}
|E_x| > \left|\frac{v_y B_z}{c}\right|
\label{eq:Emax} 
\end{equation} 
over some interval of~$x$ [see, e.g., Sagdeev \& Shapiro (1973),
Katsouleas \& Dawson (1983), Ucer \& Shapiro (2001)].  In order for
particles with~$|v_y| \sim c$ to surf, $|E_x|$ must locally
exceed~$B_z$, a condition that may arise at large Alfv\'enic Mach
number (Hoshino \& Shimada 2002).

\begin{figure}[h]
\vspace{9cm}
\includegraphics{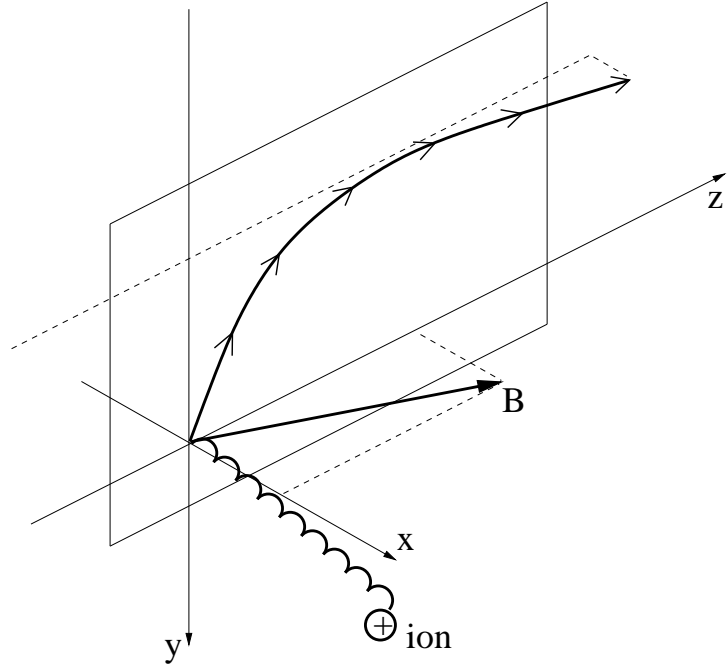}
\caption{A reproduction of figure 4 of Lee et~al (1996), which
illustrates how surfing ions escape to the upstream region 
in an oblique shock.
\label{fig:f3}}
\end{figure}

The escape of a surfing particle from the effective potential well
determines how far the particle can travel along the $-\hat{{\bf y}}$ direction,
which in turn determines the amount of energy the particle gains.
Several escape mechanisms have been considered in the literature.
First, if the maximum value of~$E_x$, denoted $E_{x,\;\rm max}$,
is less than $B_z$ (assumed constant) and a surfing particle's $|v_y|$
increases to the point that $|v_y| > c E_{x, \;\rm max}/B_z$, then the
local minimum of~$U_{\rm eff}$ disappears and the particle escapes
downstream (Sagdeev \& Shapiro 1973). 
Second, as a non-relativistic
particle's~$|v_y|$ increases, the slope of~$U_{\rm eff}$ at large~$x$
increases as illustrated in figure~\ref{fig:f2}, causing the width~$\Delta x$
of a particle's bounce oscillations to decrease. Since
$J \sim \Delta x\, \overline{p}_x$ is approximately conserved,
where~$\overline{p}_x$ is a typical value of~$|p_x|$, decreasing~$\Delta x$
increases~$\overline{p}_x$, thereby increasing~$w$~(Ucer \& Shapiro 2001).
If~$w$ increases above~$w_{\rm max}$, 
then the particle escapes downstream (Lee et~al 1996, Ucer \& Shapiro
2001). Third, in an oblique shock the velocity vector of a surfing
particle gradually rotates in the~$yz$-plane due to the
Lorentz force associated with~$B_x$, as depicted in
figure~\ref{fig:f3}, which is an adaptation of figure~4 of Lee et~al
(1996). As this happens, $|v_y|$  is eventually reduced to
zero and the particle escapes upstream since the ~$v_y B_z /c$
component of the Lorentz force ceases to turn the particle back
towards the shock (Lee et~al 1996).  Fourth, wave-particle
interactions can reduce a surfing particle's~$|v_y|$ to zero, again allowing
the particle to escape upstream (Shapiro, Lee, \& Quest 2001).  Fifth,
as noted by Lee, Shapiro, \& Sagdeev (1996) and Hoshino \& Shimada (2002),
some shocks are intrinsically nonstationary and can periodically ``break,''
allowing surfing particles to escape downstream.

For perpendicular shocks and relativistic particles with~$|v_y| \sim c$, the first 
three mechanisms described above are absent. 
This has led previous authors to consider scenarios
of ``unlimited acceleration'' when $E_x > B_z$ over
some interval in~$x$
(Katsouleas \& Dawson 1983, Ucer \& Shapiro 2001). 

\begin{figure}[h]
\vspace{9cm}
\includegraphics{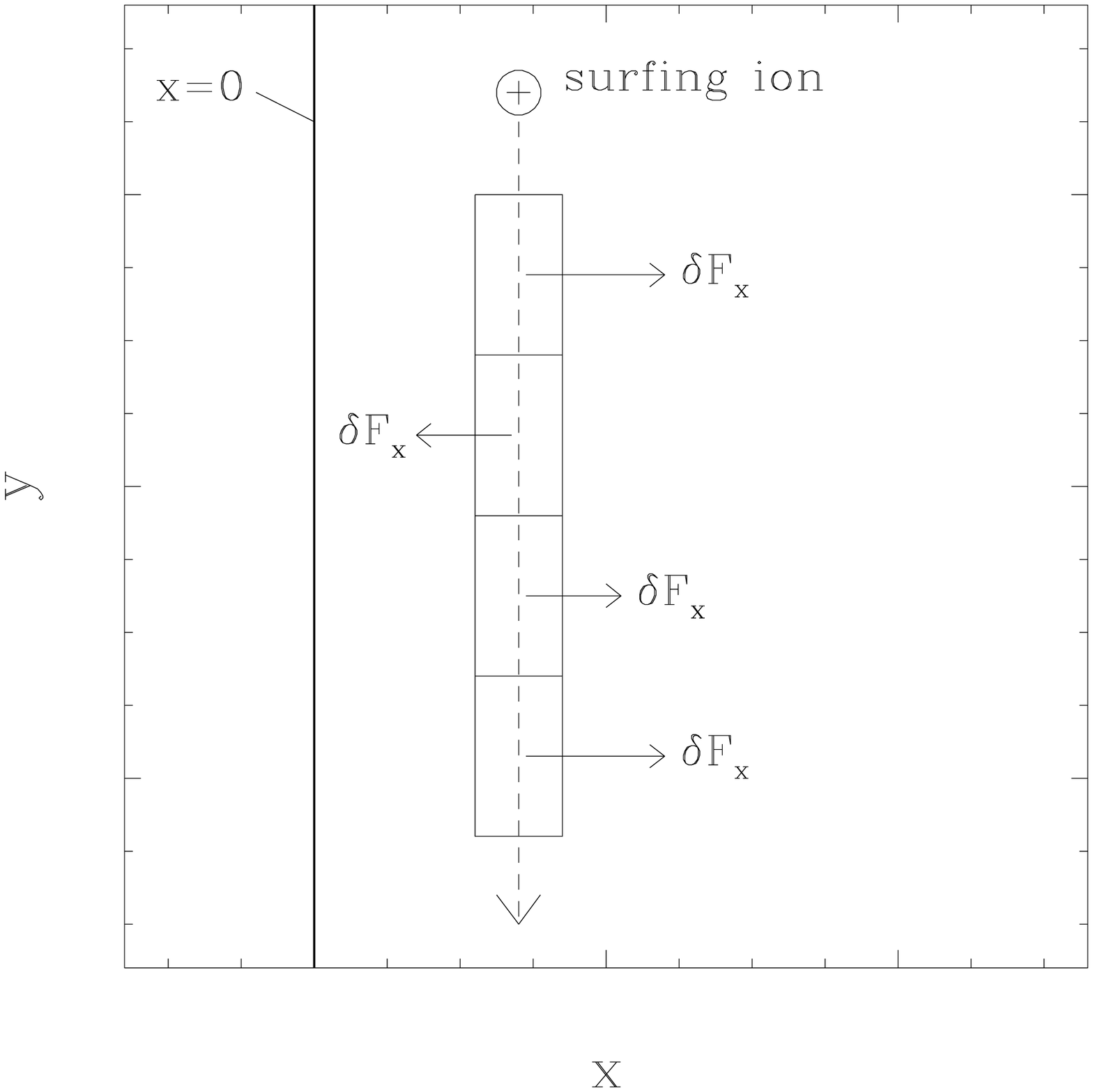}
\caption{Simple visualization of the 
origin of diffusion in~$p_x$.
\label{fig:f4}}
\end{figure}

The purpose of this paper is to show that shock surfing of
relativistic particles at perpendicular shocks leads to a power-law
energy spectrum of accelerated particles, provided fluctuations in the
shock electric and magnetic fields are not too large, and to calculate
the power-law index of this spectrum.  We ignore diffusion in a
particle's $y$-momentum, $p_y$, and assume that changes in~$p_y$ are
dominated by the motional electric field.  However, we take into
account moderate fluctuations in the electric and magnetic fields in
the shock's vicinity, and in particular the effects of these
fluctuations on the~$x$ component of a particle's momentum,~$p_x$.  An
ion propagating primarily in the~$-\hat{{\bf y}}$ direction encounters a series of
random forces in the $x$ direction (denoted~$\delta F_x$) due to
fluctuations in~$E_x$ and $B_z$, as illustrated in
figure~\ref{fig:f4}. These forces induce a series of random increments
in~$p_x$, causing diffusion in~$p_x$.
For example, suppose the typical electric and/or magnetic field
structures have a length~$l_y$ along the~$y$ direction and a typical
amplitude~$\delta E_x$, that an ion moves primarily in
the~$-\hat{\bf y}$ direction at a speed~$\sim c$,
and that these structures fluctuate on a time scale that
is longer than the time 
\begin{equation}
\triangle t \sim \frac{l_y}{c}
\label{eq:defdetlt} 
\end{equation} 
for a surfing particle to traverse one of the structures.
The random increment in~$p_x$ when
an ion moves through one structure then has a magnitude
\begin{equation}
\triangle p_x \sim \triangle t q \delta E_x.
\label{eq:defdpx} 
\end{equation} 
If the field structures are space filling, the ion undergoes
one such momentum kick during each time~$\Delta t$, leading
to a coefficient of diffusion in~$p_x$,
\begin{equation}
D_{p_x} \sim \frac{q^2 \delta E_x^2 l_y}{c},
\label{eq:dpxest} 
\end{equation} 
that is independent of particle energy.  The energy spectrum of
accelerated particles can be understood in terms of two competing
effects. On the one hand, diffusion in~$p_x$ stochastically de-traps
particles by increasing~$w$ above~$w_{\rm max}$, thereby allowing
particles to escape downstream. On the other hand, the gradual
increase in~$\gamma$ due to acceleration by~$E_y$ focuses a surfing
particle with~$|v_y| \sim c$ into the bottom of the effective
potential well, as follows from conservation of~$J$ (Ucer \& Shapiro
2001).  This combination of factors leads to a power-law energy spectrum
of accelerated particles, $dN/d\gamma \propto \gamma^{-b}$. The
value of~$b$ depends upon 
\begin{equation}
\alpha = \frac{2 D_{p_x}c}{q|E_y| w_{\rm max}} \sim \frac{t_{\rm acc}}{t_{\rm diff}},
\label{eq:intro_alpha} 
\end{equation} 
where~$t_{\rm acc} = mc\gamma/q|E_y|$ is the time to double a surfing
particle's energy and~$t_{\rm diff} = m\gamma w_{\rm max}/D_{p_x}$ is
the momentum diffusion time scale. As~$\alpha$ is decreased, the
acceleration time is reduced relative to the momentum diffusion time
scale, particles experience larger energy increases before escaping
the surfing mechanism, and one expects the energy
spectrum~$dN/d\gamma$ to become harder.  This expectation is born out
by the calculations and simulations to be presented in
sections~\ref{sec:analytic} and~\ref{sec:num}, which show that in 
the strong-scattering limit ($\alpha \gg 1$) $b\simeq \alpha$,
in the moderate-scattering limit ($\alpha \simeq 1$) $b\simeq 1-3$, and in the
weak-scattering limit ($\alpha \ll 1$)~$b \simeq 1$.

If, contrary to what we assume, the field fluctuations were so
large that the minimum of the effective potential energy disappeared
at locations separated by a typical distance~$\Delta y$ along a
surfing particle's trajectory, one would not obtain a simple power-law
energy spectrum; the spectrum would instead steepen above an
energy~$\sim q|E_y|\Delta y$ corresponding to the energy a particle
gains by surfing a distance~$\Delta y$ (Hoshino \& Shimada 2002).

As mentioned previously, if a particle is assumed to undergo diffusion
in~$p_y$ with diffusion coefficient~$D_{p_y}$ as well as ``advection''
in~$p_y$ due to~$E_y$, then a particle can in principle diffuse
to~$p_y=0$ and escape into the upstream region.  The condition under
which this escape mechanism can be neglected can be understood by
considering an analogy to first-order Fermi acceleration at a
non-relativistic shock, as follows.  Suppose a particle is a
distance~$x$ downstream of a shock, that the downstream fluid
velocity is~$u_x$, and that the particle diffuses in space with
diffusion coefficient~$D_x$. The time for a particle to return to the
shock in the absence of advection is~$t_{\rm diff,x}\sim x^2/D_x$, and
the time for the particle's distance from the shock to be doubled by
advection in the absence of diffusion is~$t_{\rm adv} = x/u_x$. The
probability that the particle returns to the shock is~$e^{-t_{\rm
diff,x}/t_{\rm adv}} = e^{-xu_x/D_x}$ (see, e.g., Bicout 1997).  This advection-diffusion
problem is analogous to 
the advection and diffusion  in~$p_y$ that would occur in shock
surfing with nonzero~$D_{p_y}$. The
probability that a particle escapes the surfing mechanism by diffusing
in momentum space to~$p_y=0$ is again $\sim e^{-t_{{\rm diff},p_y}
/t_{{\rm adv},p_y}}$, where $t_{{\rm diff},p_y} = p_y^2/D_{p_y}$ is the
time to diffuse to~$p_y=0$ in the absence of advection in~$p_y$,
and~$t_{{\rm adv},p_y}= t_{\rm acc} = mc\gamma/q|E_y|$ is the time for a
particle's $y$-momentum to double due to~$E_y$.
Escape through~$p_y$ diffusion is
thus negligible provided~$t_{\rm acc} \ll t_{{\rm diff},p_y}$, or $D_{p_y} \ll p_y^2 q
|E_y|/mc\gamma \sim q|p_y E_y|$.

We note that it is not clear how efficiently particles are injected
into the surfing acceleration mechanism. Injection appears to be
natural for pickup ions accelerated by the heliospheric termination
shock.  In the frame of the solar wind, pickup ions are born with
particle speeds comparable to the solar-wind velocity, and thus pickup
ions that arrive at the heliospheric termination shock at the right
point in their gyromotion have very small $|v_x|$, which allows them
to be trapped in front of the shock (Lee et~al 1996, Zank et~al 1996,
Lever et~al 2001).  Surfing acceleration in other cases, however, may
be more difficult to initiate. It may be that particles can be
injected at the time that a shock is formed if the shock originates
from an explosive event.  It is also possible that turbulent
conditions in the shock transition region can feed particles into the
surfing mechanism. Hoshino \& Shimada (2002) have found electron
surfing in PIC simulations of perpendicular shocks, demonstrating that
injection does occur, at least for electrons. Additional work on injection,
however, is needed.

Although our analysis is restricted to the case of perpendicular shocks,
some of the results may apply to the more general case of superluminal
shocks, for which the point of intersection between a
field line and the shock plane moves faster than~$c$. For such shocks,
there exists a ``perpendicular shock frame'' in which the shock
is stationary and the upstream electric and magnetic fields lie
in the shock plane (Begelman \& Kirk 1990). There are, however,
two issues that would have to be addressed before our results
could be generalized to this case. First, the condition for
$E_x$ to exceed~$B_z$ in more general 
superluminal shocks must be found.
Second, in general the upstream flow velocity in the perpendicular
shock frame has a component along the upstream magnetic field.
If particles start to surf with a significant~$z$ velocity,
they would only approach the state that we analyze with~${\bf v} \propto -\hat{{\bf y}}$
after significant acceleration in the~$-\hat{{\bf y}}$ direction.
As a particle approaches this
state, $|v_y|$ gradually increases, and the effects of such a gradual
increase are not included in our analysis.

The remainder of this paper is organized as follows. In
section~\ref{sec:analytic} we show analytically that surfing ions that
diffuse in~$p_x$ are described by a power-law energy spectrum. We
calculate analytically the power-law index for the case of weak
momentum diffusion ($\alpha \ll 1$), and derive a rough estimate 
for the power-law index in the case of strong momentum
diffusion ($\alpha \gg 1$). In section~\ref{sec:num} we present Monte Carlo numerical
calculations of surfing in the presence of diffusion in~$p_x$ for the
weak, moderate, and strong momentum-diffusion cases. We summarize
our results in section~\ref{sec:disc}.

\section{Analytic calculation of energy spectrum of accelerated ions}
\label{sec:analytic} 

We begin with the relativistic Vlasov equation with a term
added to model diffusion in~$p_x$,
\begin{equation}
\frac{\partial f}{\partial t}\; +\; {\bf v} \cdot \nabla f
\;+ \;q\left({\bf E} + \frac{{\bf v} \times {\bf B} }{c}\right)\cdot \nabla_{{\bf p} }f
\;\; = \;\;D_{p_x}\frac{\partial^2  f}{\partial p_x^2},
\label{eq:vl1} 
\end{equation} 
where $D_{p_x}$ is taken to be
independent of particle energy as in equation~(\ref{eq:dpxest}).
We consider a perpendicular shock and ignore spatial variations in the
magnetic field and motional electric field:
\begin{equation}
{\bf B} = B_z \hat{{\bf z}},
\end{equation} 
\begin{equation}
B_z = \mbox{ constant} > 0,
\end{equation} 
and
\begin{equation}
E_y = \mbox{ constant} < 0.
\end{equation}
We also take
\begin{equation}
E_z = 0,
\end{equation} 
\begin{equation}
E_x= - \frac{d\Phi}{dx},
\end{equation}
and
\begin{equation}
c^{-1}\frac{\partial \Phi}{\partial t} =
\frac{\partial \Phi}{\partial y} =
\frac{\partial \Phi}{\partial z} = 0.
\end{equation} 
We confine our analysis to relativistic surfing ions, for which
we assume 
\begin{equation}
v_y \simeq -c,
\end{equation} 
\begin{equation}
|v_x| \ll c, 
\end{equation} 
\begin{equation}
v_z = 0,
\end{equation} 
\begin{equation} 
\frac{\partial f}{\partial y} = \frac{\partial f}{\partial z} = 0,
\end{equation} 
\begin{equation}
\left|\frac{v_x B_z}{c}\right| \ll |E_y|,
\end{equation}
and 
\begin{equation}
\gamma = \frac{|p_y|}{mc}.
\label{eq:valgam} 
\end{equation}
Equation~(\ref{eq:vl1}) can
then be rewritten as
\begin{equation}
\frac{\partial f}{\partial t} \;+ \;v_x \frac{\partial f}{\partial x}
\;+ \;q(E_x - B_z)\frac{\partial f}{\partial p_x} \;+\; q E_y \frac{\partial f}{\partial p_y}
\;\;= \;\;D_{p_x} \frac{\partial^2 f}{\partial p_x^2}.
\label{eq:vl2} 
\end{equation} 

We now change variables, defining
\begin{equation} 
t^\prime = t,
\label{eq:deftp}  
\end{equation} 
\begin{equation}
x^\prime = x,
\label{eq:} 
\end{equation}
\begin{equation}
p_{y0} = p_y - qE_y t,
\label{eq:defpy0} 
\end{equation}  
and
\begin{equation}
w = \frac{m\gamma v_x^2}{2} + U_{\rm eff}(x) - U_{\rm min},
\end{equation}
or, equivalently,
\begin{equation}
w = -\frac{p_x^2 c}{2 p_y}  + U_{\rm eff}(x) - U_{\rm min},
\label{eq:defw} 
\end{equation} 
where
\begin{equation}
U_{\rm eff}(x) = q(\Phi + xB_z)
\label{eq:defU} 
\end{equation} 
is the effective potential energy for the particle's motion in~$x$
illustrated in figure~\ref{fig:f2},
$w$ is the energy associated with this motion,
and $U_{\rm min}$ is the value of~$U_{\rm eff}$ at its local minimum
at~$x=x_{\rm min}$. As discussed in the introduction, surfing 
particles satisfy $0 < w < w_{\rm max}$, where
\begin{equation}
w_{\rm max} = U_{\rm max} - U_{\rm min},
\label{eq:defwmax} 
\end{equation}
and $U_{\rm max}$ is the value of $U_{\rm eff}$ at its local
maximum. In terms of $(t^\prime, x^\prime, w, p_{y0})$,
equation~(\ref{eq:vl2}) becomes
\begin{equation}
\frac{\partial f^\pm}{\partial t^\prime} \;\pm\; |v_x| \frac{\partial f^\pm}{\partial x^\prime}
\;- \;\frac{q|E_y| v_x^2}{2c} \frac{\partial f^\pm}{\partial w}
\;\;= \;\;D_{p_x} v_x \frac{\partial}{\partial w}
\left(v_x \frac{\partial f^\pm}{\partial w}\right),
\label{eq:vl3} 
\end{equation} 
where $f^+$ ($f^-$) is the distribution function for particles
with~$v_x >0$ ($<0$).

We assume that the time~$\tau_B$ for a particle to bounce in the
effective potential well is much shorter than the acceleration
time or the time for a particle to escape from the potential well
due to diffusion in~$p_x$. 
We thus take 
\begin{equation} 
v_x \frac{\partial f^\pm}{\partial x^\prime}
\sim \frac{f^\pm}{\tau_B},
\label{eq:taub1}
\end{equation} 
and 
\begin{equation}
\frac{\partial f^\pm }{\partial t^\prime} \sim 
 \frac{q|E_y| v_x^2}{2c}\frac{\partial f^\pm}{\partial w}\sim
D_{p_x} v_x \frac{\partial}{\partial w}
\left(v_x \frac{\partial f^\pm}{\partial w}\right) \sim \frac{f^\pm}{\tau_0},
\label{eq:tau0} 
\end{equation} 
with 
\begin{equation}
\epsilon = \frac{\tau_B}{\tau_0}\ll 1.
\label{eq:defeps} 
\end{equation} 
We then set
\begin{equation}
f^\pm = f^\pm _0 + \epsilon f^\pm_1 + \dots
\label{eq:exp} 
\end{equation} 
and expand equation~(\ref{eq:vl3}) in powers of~$\epsilon$.
Upon collecting all terms of order~$\epsilon^0 f_0/\tau_B$ (there is only one)
and dividing by~$|v_x|$, we find that
\begin{equation}
\frac{\partial f^\pm_0}{\partial x^\prime} = 0.
\label{eq:eps0} 
\end{equation} 
As in the introduction, we take
 $x_L(w)$ and $x_R(w)$ to be the bounce points of a particle
of ``energy''~$w$, with $x_L < x_R$. Since $f^+(x_L(w),w)$ and $f^-(x_L(w),w)$
give the distribution function at the same point in phase space,
one has $f^+(x_L(w),w)) = f^-(x_L(w),w)$. Similarly, $f^+(x_R(w),w) =
f^-(x_R(w),w)$. Equation~(\ref{eq:eps0}) thus implies that
\begin{equation}
f^+_0 = f^-_0 \equiv f_0.
\label{eq:f0} 
\end{equation} 
Upon collecting all terms in equation~(\ref{eq:vl3}) of order~$\epsilon f_0/\tau_B$
and dividing by~$|v_x|$, we find that
\begin{equation}
\frac{1}{|v_x|}\frac{\partial f_0}{\partial t^\prime} \;\pm\; \frac{\partial f^\pm _1}{
\partial x^\prime} \;- \;\frac{q|E_y||v_x|}{2c}\frac{\partial f_0}{\partial w}\;\;
= \;\;D_{p_x} \frac{\partial }{\partial w} \left(|v_x|\frac{\partial f_0}{\partial w}\right).
\label{eq:eps1} 
\end{equation} 
We integrate equation~(\ref{eq:eps1}) from~$x_L(w)$
to $x_R(w)$  for $f^+$, then integrate equation~(\ref{eq:eps1}) from
$x_L(w)$ to $x_R(w)$ for $f^-$, and then add the two resulting
equations to annihilate the terms involving~$f_1^\pm$, thereby obtaining
\begin{equation}
\tau_B \frac{\partial f_0}{\partial t^\prime} \;
- \;\frac{q |E_y| g}{c \sqrt{2m\gamma}}\; \frac{\partial f_0}
{\partial w} \;\;= \;\;D_{p_x}\frac{\partial }{\partial w}
\left( g\sqrt{\frac{2}{m\gamma}} \;\frac{\partial f_0}{\partial w}\right),
\label{eq:solv} 
\end{equation} 
where
\begin{equation}
g=
\int_{x_L(w)}^{x_R(w)} dx\;\sqrt{w - U_{\rm eff}(x)+ U_{\rm min}},
\label{eq:defg} 
\end{equation} 
and 
\begin{equation}
\tau_B = \int_{x_L(w)}^{x_R(w)} \frac{dx}{|v_x|} = \sqrt{2m\gamma} \;\frac{dg}{dw}.
\label{eq:taub2} 
\end{equation}

We note the behavior of~$g(w)$ near~$w=0$ for future reference.
Near~$x=x_{\rm min}$,
\begin{equation}
U_{\rm eff}(x) = U_{\rm min} + a (x-x_{\rm min})^2 + \beta(x-x_{\rm min})^3 + \kappa(x-x_{\rm min})^4 +\dots
\label{eq:Uexp} 
\end{equation} 
Using the technique of asymptotic matching, one can show that as~$w\rightarrow 0$,
$g(w)$ has the asymptotic expansion
\begin{equation}
g(w) = \frac{\pi w}{2\sqrt{a}}\left[
1 + w\left(\frac{15 \beta^2}{32 a^3} - \frac{3\kappa}{8 a^2}\right) + \dots \right].
\label{eq:gexp} 
\end{equation} 
Since $dg/dw$ exists
and is integrable, the asymptotic expansion of~$dg/dw$ near~$w=0$ is given by
term-wise differentiation of equation~(\ref{eq:gexp}):
\begin{equation}
\frac{dg}{dw} = 
 \frac{\pi }{2\sqrt{a}}\left[
1 + 2w\left(\frac{15 \beta^2}{32 a^3} - \frac{3\kappa}{8 a^2}\right) + \dots \right].
\label{eq:dgdwexp} 
\end{equation}

We now recast equation~(\ref{eq:solv}) into the standard Fokker-Planck
form for the number of particles per unit~$w$ per unit~$p_{y_0}$,
\begin{equation}
P = \int_{x_L(w)}^{x_R(w)} \frac{(f^+ +f^-)\; dx}{|v_x|}
\simeq 2 \tau_B f_0.
\label{eq:defP} 
\end{equation} 
We start by noting from 
equations~(\ref{eq:valgam}), (\ref{eq:deftp}) and~(\ref{eq:defpy0}) that
\begin{equation}
\gamma = \frac{-p_{y0} - qE_y t^\prime}{mc},
\label{eq:valgam2} 
\end{equation} 
where both~$p_{y0}$ and~$E_y$ are negative. 
Equations~(\ref{eq:taub2}) and~(\ref{eq:valgam2}) give
\begin{equation}
\frac{\partial \tau_B}{\partial t^\prime} = \frac{\tau_B q|E_y|}{2\gamma mc}.
\label{eq:id1}
\end{equation} 
Next, we define
\begin{equation}
h = \frac{w}{g}\frac{d g}{d w},
\label{eq:defh} 
\end{equation} 
and note that
\begin{equation}
g\frac{\partial }{\partial w}\left(\frac{P}{2\tau_B}\right) = \frac{1}{2\sqrt{2m\gamma}}
\left[
\frac{\partial }{\partial w}\left(\frac{Pw}{h}\right) - P\right].
\label{eq:id2} 
\end{equation} 
Throughout the rest of this section we drop the primes, setting~$t^\prime\rightarrow t$.
Substituting $f_0 = P/2\tau_B$ into equation~(\ref{eq:solv}) and
making use of equations~(\ref{eq:id1}) and (\ref{eq:id2}),
we find
\begin{equation}
\frac{\partial P}{\partial t}=  - \frac{\partial \Gamma}{\partial w},
\label{eq:fp1} 
\end{equation} 
where
\begin{equation}
\Gamma = 
\left( \frac{D_{p_x}}{m\gamma} - \frac{q|E_y| w}{2mc\gamma h} \right)P
- \frac{\partial}{\partial w}\left(\frac{D_{p_x}w P}{m\gamma h}\right)
\label{eq:defGamma} 
\end{equation} 
is the flux of particles in~$w$-space. The  quantity in
the first set of parentheses on the right-hand side of equation~(\ref{eq:defGamma})
is~$\langle \triangle w\rangle/\triangle t$, while the coefficient
of~$P$ within the second set of
parentheses is~$\langle (\triangle w)^2\rangle/2\triangle t$, where
$\triangle w$ is the change in~$w$ during a small time
interval~$\triangle t$, and the angle brackets denote an average
over the stochastic scattering that gives rise to diffusion
in~$p_x$. The term~$-q|E_y| w/2mc\gamma h$ is the rate
of change of~$w$ caused by the approximate conservation of the
adiabatic invariant
\begin{equation}
J = \int_{x_L(w)}^{x_R(w)} p_x \;dx = g\sqrt{2 m\gamma} .
\label{eq:defJ} 
\end{equation} 
That is, setting $dJ/dt=0$ yields $dw/dt = -q|E_y| w/2mc\gamma h$,
which confirms the statement made in the introduction that
$J$-conservation drives a relativistic surfing particle
towards the
bottom of the effective potential well (Ucer \& Shapiro 2001). The
terms on the right-hand side of equation~(\ref{eq:defGamma})
proportional to~$D_{p_x}$ give the advection and diffusion
in~$w$-space arising from a time average of a particle's diffusion
in~$p_x$.

If diffusion in~$p_x$ causes a particle's~$w$ to increase to~$w_{\rm max}$,
the particle escapes downstream. 
We thus seek a solution to equation~(\ref{eq:fp1}) subject to the boundary
condition
\begin{equation}
P(w_{\rm max},t) = 0.
\end{equation} 
We also require that~$P$ be finite and differentiable at~$w = 0$.
Since there are no $p_{y0}$ derivatives in equation~(\ref{eq:fp1}),
we can treat particles with different values of~$p_{y0}$
independently. We will henceforth focus on particles with a single
value of~$p_{y0}$, and assume that a large number of such particles
are injected into 
the surfing mechanism at~$t=0$. We rewrite equation~(\ref{eq:valgam2}),
dropping the  prime on~$t^\prime$, to obtain
\begin{equation}
\gamma = \left(\frac{q|E_y|}{mc}\right) (d + t),
\label{eq:gam1} 
\end{equation} 
where $d = p_{y0}/qE_y$ is the time required to double the
particle's initial $y$-momentum.
Because a surfing
particle's energy increases linearly in time,
the number of particles that are accelerated to a final Lorentz
factor in the interval~$(\gamma, \gamma+d\gamma)$,
denoted~$\displaystyle \frac{dN}{d\gamma} d\gamma$, is equal to the
number of particles that escape the effective potential well in the
time interval~$(t,t+dt)$, where $t = (mc\gamma/q|E_y|) - d$ and $dt =
(mc/q|E_y|)d\gamma$ from equation~(\ref{eq:gam1}).  Thus,
\begin{equation}
\frac{dN}{d\gamma} =  \frac{mc \Gamma(w_{\rm max},t)}{q|E_y|} \left|
\begin{array}{l}
\\
_{t= (mc\gamma/q|E_y|) - d}\\
\end{array}\right. .
\label{eq:dNdg} 
\end{equation} 

Using equation~(\ref{eq:gam1}), we can rewrite equation~(\ref{eq:fp1}) as
\begin{equation}
(d+t) \frac{\partial P}{\partial t}
= \frac{\partial }{\partial u} \left[
\left(\frac{u}{2h} - \frac{\alpha}{2}\right) P\right]
+ \frac{\partial ^2}{\partial u^2} \left(
\frac{\alpha u P}{2h}\right) \equiv{\cal L}(P)
\label{eq:fp2} 
\end{equation} 
where 
\begin{equation}
u =\frac{w}{w_{\rm max}},
\label{eq:defu} 
\end{equation} 
and 
\begin{equation}
\alpha = \frac{2 D_{p_x}c}{q|E_y| w_{\rm max}}.
\label{eq:defalpha} 
\end{equation} 
The eigenvalue equation
\begin{equation}
{\cal L}(P_n) = - \lambda_n P_n
\label{eq:eig} 
\end{equation} 
can be written in Sturm-Liouville form. 
We consider eigenfunctions that
are finite and differentiable at~$u=0$ and satisfy
\begin{equation}
P_n(1) = 0.
\label{eq:Pn1} 
\end{equation} 
For these boundary conditions,
and since $g(0) = 0$, the eigenvalues
are real and the eigenfunctions
of~${\cal L}$ form a complete set on the interval $0<u<1$.
The solution for~$P(u,t)$ can thus be expanded as
\begin{equation}
P(u,t) = \sum_{n=0}^{\infty} \chi_n(t) P_n(u).
\label{eq:expP} 
\end{equation} 
From equation~(\ref{eq:fp2}),
\begin{equation}
\chi_n(t) = \chi_{n}(0) \left(\frac{d}{d+t}\right)^{\lambda_n},
\label{eq:cn} 
\end{equation} 
where $\chi_{n}(0)$ is determined from the initial
conditions on~$P$. 
We order the eigenvalues and their corresponding eigenfunctions
so that  $\lambda_0 < \lambda_1 < \lambda_2 < \dots,$~etc. 
For $t\gg d$, $P(u,t)$ is dominated by the first
term in the sum in equation~(\ref{eq:expP}),
\begin{equation}
P(u,t) \propto  P_0(u)t^{-\lambda_0},
\label{eq:longtime} 
\end{equation} 
and, from equations~(\ref{eq:defGamma}), (\ref{eq:dNdg}), and~(\ref{eq:longtime}),
\begin{equation}
\frac{dN}{d\gamma} \propto \gamma^{-b},
\label{eq:lt2} 
\end{equation} 
where
\begin{equation}
b = 1 + \lambda_0.
\label{eq:defb} 
\end{equation} 

We note that
\begin{equation}
{\cal L}(P_n ) = \frac{\alpha}{2} \frac{\partial}{\partial u}
\left[
 e^{-u/\alpha}g \frac{\partial}{\partial u}
\left( \frac{e^{u/\alpha}P_n}{g^\prime}\right)\right],
\label{eq:L} 
\end{equation}
where
\begin{equation}
g^\prime = \frac{dg}{du}.
\label{eq:gp} 
\end{equation} 
Multiplying equation~(\ref{eq:eig})  by $e^{u/\alpha} P_n/g^\prime$, integrating
from~$u=0$ to~$u=1$, and integrating by parts, we find
that
\begin{equation}
\lambda_n > 0.
\label{eq:lgz} 
\end{equation} 

\subsection{Smallest eigenvalue in the limit of weak momentum diffusion ($\alpha \ll 1$)}
\label{eq:smallalpha} 

We now calculate~$\lambda_0$ to leading order in~$\alpha$ when~$\alpha
\ll 1$ for any~$U_{\rm eff}(x)$ with a single potential well as in
figure~\ref{fig:f2}.  We take
\begin{equation}
P_0(0) = 1.
\label{eq:Pn0} 
\end{equation} 
From Sturm-Liouville theory, we know
that~$\lambda_0$ is the unique value of~$\lambda_n$ in
equation~(\ref{eq:eig}) for which~$P_n$ satisfies the boundary
conditions and has no zeroes in the interval~$0 < u <
1$.  We use this fact to calculate~$\lambda_0$ in the following four
steps: (1) we assume an order of magnitude for~$\lambda_0$,
(2) we derive the resulting solution for~$P_0$,  (3) we
determine the value of~$\lambda_0$ by requiring that~$P_0(0)=1$
and~$P_0(1)=0$, and finally (4) we verify that~$P_0$ has no zeroes
for~$0<u<1$.

Starting with step~1, we set
\begin{equation}
\lambda_0 = \alpha^{-1} e^{-1/\alpha} \theta
\label{eq:deftheta} 
\end{equation} 
and assume that~$\theta$ is of order unity.  This value is close to
$\lambda$'s lower limit of~0.  The corresponding  particle energy
spectrum is hard ($b\simeq 1$), which is expected in the small-$\alpha$ limit, since the
acceleration time is much less than the momentum-diffusion time [see
equation~(\ref{eq:intro_alpha})].

We now carry out step 2. We  set
\begin{equation}
P_0 = \frac{g^\prime}{g^\prime(0)} \left[e^{-u/\alpha} + e^{-1/\alpha} F\right],
\label{eq:defF} 
\end{equation} 
and rewrite~${\cal L}(P_0) = - \lambda_0 P_0$ as
\begin{equation}
 \frac{\partial }{\partial u} \left[
e^{-u/\alpha} g \frac{\partial }{\partial u} \left(
e^{u/\alpha}F\right)\right]
= - \frac{2\theta g^\prime}{\alpha^2} \left(
e^{-u/\alpha} + e^{-1/\alpha}F\right) .
\label{eq:FH} 
\end{equation} 
Equations~(\ref{eq:Pn0}) and (\ref{eq:Pn1}) give
\begin{equation}
F(0) = 0,
\label{eq:Fu0} 
\end{equation} 
and
\begin{equation}
F(1) =-1.
\label{eq:Fu1} 
\end{equation} 
We assume that~$F$ decreases monotonically
from~0 to~$-1$ as $u$ increases from~0 to~1, an assumption
that we verify at the end. Then for
$|1-u| \gg \alpha$, we can take 
$e^{-u/\alpha} + e^{-1/\alpha}F \simeq e^{-u/\alpha}$
on the right-hand side of equation~(\ref{eq:FH}).
In addition, when $ |1-u| \ll 1$, the entire right-hand side
of equation~(\ref{eq:FH})  is of order~$e^{-1/\alpha}$
and can be replaced by~0 to an excellent approximation.
Thus, throughout the interval~$0<u<1$, we can neglect
the second term in parentheses on the right-hand side
of equation~(\ref{eq:FH}), and write
\begin{equation}
 \frac{\partial }{\partial u} \left[
e^{-u/\alpha} g \frac{\partial }{\partial u} \left(
e^{u/\alpha}F\right)\right]
= -\frac{2\theta g^\prime e^{-u/\alpha}}{\alpha^2} .
\label{eq:FH2} 
\end{equation} 
Integrating twice, using $g(0)  = 0$ from equation~(\ref{eq:gexp}),
and imposing equation~(\ref{eq:Fu0}), we find that
\begin{equation}
F = - \frac{2\theta e^{-u/\alpha} }{\alpha^2}
\int_0^u du_1\;\frac{ e^{u_1/\alpha}}{g(u_1)} \int_0^{u_1}
du_2 \;e^{-u_2/\alpha}g^\prime(u_2).
\label{eq:Fouter} 
\end{equation} 

We now carry out step~3. Imposing equation~(\ref{eq:Fu1}) and evaluating the
integral in equation~(\ref{eq:Fouter}) to lowest order
in~$\alpha$, we find
\begin{equation}
\theta =\frac{g(1)}{2g^\prime(0)}.
\label{eq:theta} 
\end{equation} 

For step 4, we note that for~$0<u<1$, $g$ and~$g^\prime$ are positive.
Thus, $F$ decreases monotonically from~$0$ to~$-1$ as~$u$ is increased
from~0 to~1, as assumed.  This implies that~$P$ has no zeroes
for~$0<u<1$, and that the value of~$\lambda_0$ given by
equations~(\ref{eq:deftheta}) and (\ref{eq:theta}) is indeed (a good
approximation of) the smallest eigenvalue.

Equations~(\ref{eq:dgdwexp}), (\ref{eq:deftheta}), and~(\ref{eq:theta}) give
\begin{equation}
\lambda_0 = \frac{\alpha^{-1} e^{-1/\alpha}g(1) \sqrt{a}}{\pi w_{\rm max}},
\label{eq:lambda0} 
\end{equation} 
where~$a$ is defined by equation~(\ref{eq:Uexp}).
The power-law index of the energy spectrum of accelerated particles
is then
\begin{equation}
b \; = \;1 \;+ \;\frac{\alpha^{-1} e^{-1/\alpha}g(1) \sqrt{a}}{\pi w_{\rm max}}.
\label{eq:bweak} 
\end{equation} 
From equation~(\ref{eq:defF}), it can be seen that~$P_0$ is strongly
peaked near~$u=0$. This means that if a large number of particles
start surfing at~$t=0$, the majority of those remaining at large times will
be concentrated near the bottom of the effective potential
well (with $u\lesssim \alpha$ since
$P_0\simeq e^{-u/\alpha} \mbox{ for $u\ll 1$}$).

For reference, the higher eigenvalues can be found by expanding
equation~(\ref{eq:eig}) in powers of~$\alpha$, solving for $P/h$
in an inner region with~$u \ll 1$ and an outer region with~$u \gg \alpha$,
and then matching the inner and outer solutions in the region~$\alpha \ll u \ll 1$.
One finds that $P/h$ satisfies Kummer's equation in the inner region regardless
of the precise form of $U_{\rm eff}(x)$. This is because~$h\simeq 1$
in the inner region regardless of the exact form of~$U_{\rm eff}(x)$.
The inner solution can be matched to the outer solution only if
\begin{equation}
\lambda_n \simeq \frac{n}{2}
\label{eq:psin} 
\end{equation} 
for~$n=1,2,\dots$, provided~$n$ is of order unity (e.g.,$ \ll
\alpha^{-1}$). The~$n$ zeroes of~$P_n$, denoted~$u_1, u_2, \dots , u_n$,
occur in the inner region for~$n$ of order unity, with each of the~$u_i$
of order~$\alpha$.

\subsection{Smallest eigenvalue in the limit of strong momentum diffusion ($\alpha \gg 1$)}
\label{sec:strong}

When~$\alpha \gg 1$, 
we can write equation~(\ref{eq:eig})  to lowest 
order in~$\alpha^{-1}$ as
\begin{equation}
\frac{\partial^2}{\partial u^2}\left(\frac{uP}{2h}\right)
- \frac{\partial }{\partial u}\left(\frac{P}{2}\right) = - \psi P,
\label{eq:sd} 
\end{equation} 
where
\begin{equation}
\lambda  = \alpha \psi.
\label{eq:defpsi} 
\end{equation} 
We again seek a solution for~$P(u)$ that satisfies $P(0) = 1$ and
$P(1) = 0$. In going from equation~(\ref{eq:eig}) to
equation~(\ref{eq:sd}) we have retained the highest order
derivative. We thus have a regular perturbation problem and do not
expect a boundary layer.  Since there is no small parameter in
equation~(\ref{eq:sd}), we expect the smallest value of~$\psi$ to be
of order unity. Then, to lowest order in~$\alpha^{-1}$,
\begin{equation}
b \simeq \alpha
\label{eq:bstrong} 
\end{equation} 
to within a factor of order unity.

\section{Numerical calculation of energy spectrum of accelerated particles}
\label{sec:num} 

In this section, we confirm and extend the analytic results of
section~\ref{sec:analytic}  with the use of Monte Carlo test-particle
calculations. We follow a set of test particles (ions) with~$p_z=0$ that obey the
equations of motion
\begin{equation}
\frac{dp_x}{dt}\; = \;q\left(E_x + \frac{v_y B_z}{c}\right) + \tilde{\delta p_x},
\label{eq:mcpx} 
\end{equation}  
and
\begin{equation}
\frac{dp_y}{dt} \; = \; q\left(E_y - \frac{v_x B_z}{c}\right),
\label{eq:mcpy}
\end{equation} 
where $\tilde{\delta p_x}$ is a stochastic function of time that increments~$p_x$
by~$\pm \Delta p_x$ (equal chance for~+ or~-) during each time step~$\Delta t$,
where~$\Delta p_x = \sqrt{ 2 D_{p_x} \Delta t}$ and $D_{p_x}$ is a constant. We
work in the shock wave frame, with $E_x = - d\Phi/dx$, where
\begin{equation}
\Phi(x) = \Phi_0 e^{-(x/d)^2}.
\label{eq:mcphi} 
\end{equation} 
We set $E_y=-\gamma_{sh}v_{sh}B_{0}/c$, where $v_{sh}$, $\gamma_{sh}$, and
$B_{0}$ are the shock speed, the Lorentz factor for the shock speed,
and the upstream magnetic field in the laboratory frame, respectively.
Although $B_z$ is in general not constant, we assume
a constant value, $B_z=\gamma_{sh}B_{0}$. 
We set~$v_{sh} = 0.25c$, $\gamma_{sh} = 1.0328$, and $\Phi_0/B_0\,d=4.5$.
The minimum of the approximate effective potential $\Phi + xB_z$ occurs
at~$x=1.63d$. We start the particles at~$x=2.25d$, with~$p_x = -0.1m_i c$,
$p_y = -10m_i c$, and~$p_z = 0$.
By varying~$D_{p_x}$, we are able to consider different values
of~$\alpha$ in equation~(\ref{eq:defalpha}).

To improve the statistics, we implement particle splitting as follows.
We start each simulation with~1000 particles.  At each time step, we
update $\bf p$ and $\bf x$ for each particle. When a particle escapes
from the effective potential well the calculation for that particle is
stopped and $\gamma$ at that time is recorded. When half of the
original particles have escaped, each of the remaining particles is
``split''---that is, a copy of each particle is created with the same
position and momentum, and both the original and the copy are
subsequently tracked.  When we calculate the energy spectrum, the
weightings of split particles are halved. The splitting process is
repeated each time that the particle number drops to~500 until the end
of the simulation, when every particle's $\gamma$ exceeds a threshold
value,~$\gamma_{m} = 500$.

In the weak scattering case,~$\alpha\ll 1$, a difficulty for
calculating the spectrum is that almost all of the particles reach the
threshold~$\gamma_m$ before escaping. We therefore use another method
to obtain the spectrum when~$\alpha\ll 1$. We assume that the spectrum
is given by a power law, $dN/d\gamma\propto \gamma^{-b}$
with~$b>1$. The number of particles with~$\gamma$ between $\gamma_1$
and $\gamma_2$ is given by
\begin{equation}
N_{1\rightarrow
2}=C\int^{\gamma_2}_{\gamma_1}\gamma^{-b}d\gamma=\frac{C(\gamma_2^{1-b}-\gamma_1^{1-b})}{1-b},
\label{nu1}
\end{equation}
where $C$ is a constant. The number of particles with~$\gamma>\gamma_2$
is
\begin{equation}
N_{2\rightarrow\infty}=-\frac{C \gamma_2^{1-b}}{1-b}.
\label{nu2}
\end{equation}
From Eqs. (\ref{nu1}) and (\ref{nu2}), we obtain the spectral index $b$ as
\begin{equation}
b=1-\frac{\ln \left(1+N_{1\rightarrow
2}/N_{2\rightarrow\infty}\right)}{\ln \left(\gamma_1/\gamma_2\right)}.
\end{equation}
We take $\gamma_1=250$ and $\gamma_2=500$ and 
track 5000 particles for each value of~$\alpha$.

\begin{figure}[h]
\vspace{9cm}
\includegraphics{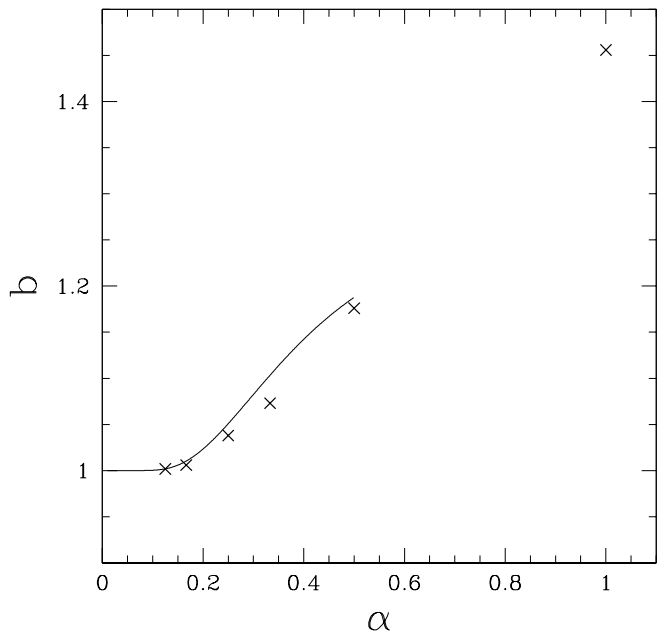}
\caption{Power-law index~$b$ of the energy spectrum of accelerated
particles as a function of the dimensionless momentum-diffusion
parameter~$\alpha$. The crosses are the results of Monte Carlo
calculations. The solid line is the analytic result
in equation~(\ref{eq:bweak}).
\label{fig:f5}}
\end{figure}

\begin{figure}[h]
\vspace{9cm}
\includegraphics{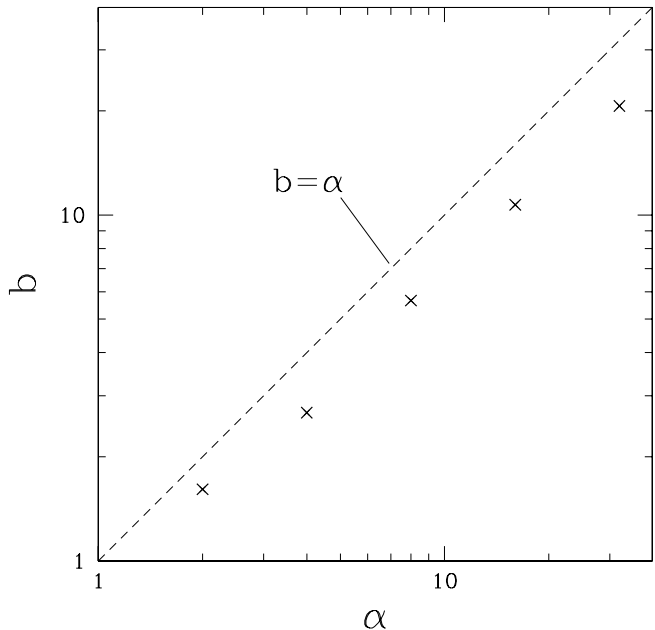}
\caption{Log-log plot of the power-law index~$b$ of the
energy spectrum of accelerated
particles as a function of the dimensionless momentum-diffusion
parameter~$\alpha$. The crosses are the results of Monte Carlo
calculations. The dashed line gives the approximate result
in equation~(\ref{eq:bstrong}).
\label{fig:f6}}
\end{figure}

Our numerical results for~$\alpha \leq 1$ are plotted in figure~\ref{fig:f5}
along with our analytic result [equation~(\ref{eq:bweak})] for the
same shock parameters.
The numerical results and analytic result converge for~$\alpha \ll 1$.
Our numerical results for $\alpha > 1$ are plotted in figure~\ref{fig:f6}, and
are consistent with equation~(\ref{eq:bstrong}) when~$\alpha \gg 1$.

\section{Summary}
\label{sec:disc} 

In this paper we model surfing acceleration of relativistic particles
at perpendicular shocks as a process in which particles undergo
diffusion in~$p_x$ ($x$ being the direction of the shock normal) while
propagating in steady-state electric and magnetic field profiles.  For
shocks with~$E_x > B_z$ over some interval in~$x$, where~$z$ is the
direction of the magnetic field, and for relativistic ions propagating
primarily in the $-\hat{\bf y}$ direction, surfing produces a
power-law energy spectrum of accelerated particles, provided the
momentum diffusion coefficient~$D_{p_x}$ is independent of energy.  We
have calculated analytically the power-law index for the case of weak
momentum diffusion, given in equation~(\ref{eq:bweak}), and have
carried out Monte Carlo test-particle calculations to determine the
power-law index for the weak, moderate, and strong momentum diffusion
cases (section~\ref{sec:num}).

\acknowledgements
We thank Eric Blackman, Steve Cowley,
and the anonymous referee for helpful comments.
This work was supported by NSF grant AST-0098086 and DOE grants
DE-FG02-01ER54658 and DE-FC02-01ER54651 at the University of Iowa.

\references

Begelman, M., \& Kirk, J. 1990, ApJ, 353, 66

Bicout, D. 1997, Phys. Rev. E., 56, 6656

Hoshino, M., \& Shimada, N. 2002, ApJ, 572, 880

Katsouleas, T., \& Dawson, J. 1983, Phys. Rev. Lett., 51, 392

Lee, M., Shapiro, V., \& Sagdeev, R. 1996, J. Geophys. Res., 101, 4777

Lever, E., Quest, K., \& Shapiro, V. 2001, Geophys. Res. Lett., 28, 1367

Lipatov, A., \& Zank, G. 1999, Phys. Rev. Lett., 82, 3609

McClements, K., Dieckmann, M., Ynnerman, A., Chapman, S., \& Dendy, R. 2001, Phys. Rev.
Lett., 87, 255002

Ohsawa, Y., \& Sakai, J. 1987, ApJ, 313, 440

Sagdeev, R. 1966, Reviews of Plasma Physics, ed. M. Leontovich
(Consultants Bureau: New York), vol. 4, p. 23

Sagdeev, R., \& Shapiro, V. 1973, JETP Lett., 17, 279

Shapiro, V., Lee, M., \& Quest, K. 2001, J. Geophys. Res., 106, 25023

Ucer, D., \& Shapiro, V. 2001, Phys. Rev. Lett., 87, 075001

Zank, G., Pauls, H., Cairns, I., \& Webb, G. 1996, J. Geophys. Res., 101, 457

\end{document}